\documentclass[preprint,12pt]{elsarticle}
\usepackage{graphicx}
\usepackage{amssymb}
\usepackage{amsmath}
\usepackage{rotating}
\usepackage{lineno}
\usepackage{float}
\usepackage{subfigure}
\newcommand{\ra}[1]{\renewcommand{\arraystretch}{#1}}
\biboptions{sort&compress, square}

\journal{ }

\begin{document}

\begin{frontmatter}



\title{Influence of model asymmetry on kinetic pathways in binary Fe-Cu alloy: a kinetic Monte Carlo study}


\author[label1,label3]{David Bombac}
\author[label1]{Goran Kugler\corref{cor1}}

\cortext[cor1]{Corresponding author}
\address[label1]{Faculty of Natural Sciences and Technology, University of Ljubljana, Askerceva 12, SI-1000 Ljubljana, Slovenia}
\address[label3]{Department of Physics, King’s College London, The Strand, London WC2R 2LS, U.K.}

\begin{abstract}
A kinetic Monte Carlo simulations with model asymmetry in binary Fe-Cu alloy leading to the same microstructure are presented. A method based on thermodynamic data for calculation of interatomic potentials dependent on model asymmetry is presented and evaluated.
Results show that kinetic pathways are sensitive to model asymmetry and are compared to the classical growth and coarsening theories. Experimental diffusion data is used and compared to simulation results to determine a realistic combination for simulations.

\end{abstract}


\end{frontmatter}


\section{Introduction}
\label{intro}

Phase transformations are the dominant mechanism for microstructure development and control in crystalline solids. From the thermodynamic point of view, specific boundary in a binary phase diagram depends only on a solubility limit~\cite{christian} which is correlated to the mixing energy of alloy elements. At specific temperature only one value of the mixing energy leads to obtained microstructure. When phase transformations are governed by diffusion process, the governing mechanism is the vacancy/atom exchange. This mechanism can be very effectively simulated by a kinetic Monte Carlo (kMC) method~\cite{young}.  

Vacancy diffusion is thermally activated mechanism with a mixing energy that governs response of the system toward clustering or phase separation. With kMC simulations, study of diffusion at a truly atomistic spatial scale is possible. With time scales long enough to determine kinetic pathways in real alloys when the model is appropriately parametrized~\cite{SoisAM96, AthAM96,schmauder,cerezo,Mao2011, BombacPhD,Warczok201259}.

When simulations are used to address behaviour in real physical alloy systems, a simple binary model allow observation of the response where only one variable is changed, e.g. concentration $c$, temperature $T$, stress $\Sigma$ or strain $\epsilon$. From positional simplicity  Fe-Cu alloy system is often used as a benchmark system for various simulation methods~\cite{Osetsky1994236,SoisAM96,schmauder,BombacPhD,Nagano2006223,Deschamps2010236}. This alloy system is important in industrial practice for its age hardening behaviour in ferritic and martensitic steels, in pressure vessel steels for radiation and high temperature embrittlement and also for efficient use of recycled steel.

The Fe-Cu alloy system has a large and almost symmetrical miscibility gap with small differences in atom sizes and has been extensively studied experimentally. Experimental  data confirms that Cu clusters with sizes up to 2 nm are fully coherent with the $\alpha$-iron matrix~\cite{othenPM, pizzini, othenPML, hrtem95} which justifies use of rigid lattice in simulations. Furthermore, availability of experimental data enables comparison and verification of results from kMC simulations with a large collection of data from physical experiments~\cite{salje1833,LBdiff,othenPM,pizzini,othenPML,hrtem95,monzenPM,apfecu,Golubov2000,Takahashi2011}.

Previous studies~\cite{FraPRB94,athenesPM97,athenesPM99,AthAM00,rousselPRB,soissonCluster2007,Warczok201259} revealed that kinetic pathways can be influenced by changes to asymmetry of atomic mobility which is in kMC simulations defined through asymmetry parameter $a^*$. Value of $a^*$ has large influence on precipitation and kinetic pathways especially in later stages of growth and during coarsening. This was studied for various binary alloys and also with conditions not corresponding to realistic diffusion behaviour~\cite{FraPRB94,athenesPM97,athenesPM99,AthAM00,rousselPRB,soissonCluster2007,Erdelyi20105639,Warczok201259}. In the majority of studies performed simulations using the energetic model based on the kinetic Ising model and the ghost energies between vacancy and its nearest neighbours were not considered, while only first nearest neighbour interaction were used. In \cite{soissonCluster2007} energetic parameters were derived on basis of density functional theory and influence of asymmetry parameter $a^*$ on diffusion and changes of kinetic pathways were not studied. 

The  current study presents a simple saddle point energy broken bond model based on cohesive energy, mixing energy and asymmetry parameter $a^*$ that can be applied also to other binary systems. Reporting values for first and second nearest neighbour interatomic interaction energies, elucidating results obtained with arbitrarily chosen values of $a^*$ in negative and positive range and comparing the obtained results to previous simulations and experimental diffusion data. Influence of vacancy trapping and its effects is explored for values of asymmetry parameter $a^*$ corresponding to the realistic precipitation behaviour of Cu in Fe-Cu alloys.

The paper is organized as follows. The diffusion model and its parametrization are detailed in Section II. Results of simulations are presented and discussed in section III. Details of simulations parameters are given in Appendix A.

\section{Diffusion model and parametrization}

Evolution of the alloy microstructure at high temperatures is governed with vacancy diffusion. In simulations presented, the model parametrization is based on experimental thermodynamic data and random walk theory. Within transition state theory the rate probability for a of vacancy-solute exchange ${\Gamma}_{\rm{XV}}$ between atom X and vacancy V is given by
\begin{equation} 
	{\Gamma}_{\rm{XV}}={{\nu}}_{\rm{X}} \; {\exp}{\left({-\frac{\Delta{E_{\rm{XV}}}}{k_{\text{B}} T}}\right)}
	\label{transrate}
\end{equation}
where $\nu_{\rm{X}}$ is the attempt frequency, $k_{\rm{B}}$ is the Boltzmann constant, $T$ is the temperature and $\Delta{E_{\rm{XV}}}$ is the energy barrier associated with the vacancy-solute exchange and is dependent on local atomic configuration. For a given configuration the sum over all possible rate probabilities, $\Gamma_{\rm{tot}}=\sum_i \Gamma_{\rm{XV},i}$, is evaluated and exchange $i$ is selected according to
\begin{equation} 
\sum_{i=0}^{n-1} \Gamma_{\rm{XV},i} < \rho_1 \Gamma_{\rm{tot}} \leq \sum_{i=0}^{n} \Gamma_{\rm{XV},i}
\end{equation} 
where $\rho_1$ is uniform random number between 0 and 1, and $n$ is the number of all possible events for one transition. The kMC simulates Poisson processes, the simulations time $t_{\rm{kMC}}$ of each exchange $\Delta t$ is evaluated as~\cite{young,Bortz75,gillespie}
\begin{equation} 
\Delta t_{\rm{kMC}} = \frac{-\ln(\rho_2)}{\Gamma_{\rm{tot}}} \approx \frac{1}{\Gamma_{\rm{tot}}}
\end{equation} 
Since vacancy concentration in simulation box exceeds real vacancy concentration, the physical time needs to be adjusted for this difference. Accounting for vacancy concentration ,the real physical time $t$ is given as
\begin{equation}
t = \frac{c_{\text{Vsim}}}{c_{\text{Veq}}} t_{\rm{kMC}}
\label{eq:11}
\end{equation}
where, $c_{\text{Vsim}}$ is the vacancy concentration in the simulation box, and $c_{\text{Veq}}$ is the equilibrium vacancy concentration in the alloy \cite{schmauder, BombacPhD, LeBou}.

The energy barrier $\Delta E_{\rm{XV}}$ is obtained within a representation of the broken bond model~\cite{BombacPhD} for each possible event. There are two contributions, the energy associated with the saddle point binding energy controlling kinetic properties, and the cohesive energy described as sum of the pair wise interactions of all the broken bonds associated with the solute-vacancy exchange  controlling equilibrium properties. $\Delta E_{\rm{XV}}$ is given by
\begin{equation} 
	\Delta E_{\rm{XV}}=e_{\rm{SP}}^{X}-\sum_{k,i} \epsilon_{{\rm{X}}i}^{(k)}-\sum_{k,j} \epsilon_{{\rm{V}}j}^{(k)}
	\label{migracijskaE}
\end{equation}
where the number $k$ represents nearest neighbour position. In our case only first and second nearest neighbour positions were accounted for. The saddle point energy $e_{\rm{SP}}$ represents energy state of the atom at the saddle-point between its initial and final positions. In the case of Fe-Cu binary system the homoatomic interaction energy for Fe $\epsilon_{{\rm{FeFe}}}$ was calculated from the cohesive energies of pure Fe given as 
\begin{equation}
E_{\text{cohFe}}=\frac{z_1}{2} \epsilon_{\text{FeFe}}^{(1)} + \frac{z_2}{2} \epsilon_{\text{FeFe}}^{(2)}
\label{par_cohesive}
\end{equation}
where $z_1$ and $z_2$ are numbers of the first and second neighbours, respectively \cite{ducastelle1991order}. Heteroatomic interaction energy $\epsilon_{{\rm{FeCu}}}$ is associated to the mixing energy of the Fe-Cu system as
\begin{equation}
E_{\text{mixFeCu}} = \frac{z_1}{2} \left( {\epsilon_{\text{FeFe}}^{(1)}} + {\epsilon_{\text{CuCu}}^{(1)}} - {2\epsilon_{\text{FeCu}}^{(1)}} \right) + \frac{z_2}{2} \left({\epsilon_{\text{FeFe}}^{(2)}} + {\epsilon_{\text{CuCu}}^{(2)}} - {2\epsilon_{\text{FeCu}}^{(2)}} \right)
\label{mixinfFeCu}
\end{equation}
Note, that phase diagram depends only on the mixing energy which was estimated from the solubility limit of Cu in Fe~\cite{salje1833}. 
In order to evaluate different kinetic pathways in the Fe-Cu binary system an asymmetry parameter $a^*$, which accounts for the  differences in diffusion of {Fe} and {Cu} was defined through the solute-solute and solute-vacancy interaction energies as
\begin{equation}
a^*= \sum_{i=1}^{k}\frac{\epsilon_{\text{FeFe}}^{(i)} - \epsilon_{\text{CuCu}}^{(i)}}{\epsilon_{\text{FeFe}}^{(i)} + \epsilon_{\text{CuCu}}^{(i)} - 2\epsilon_{\text{FeCu}}^{(i)}}
\label{asymmetry1}
\end{equation}
Differences in diffusion of each element can also be introduced through the saddle point energies, which were in our case set to be equal ($e_{\rm{SP}}=-9.25$ eV).

By reordering equations (\ref{par_cohesive}), (\ref{mixinfFeCu}), (\ref{asymmetry1}) and assumption $\epsilon_{\text{XX}}^{(2)}= 0.5 \epsilon_{\text{XX}}^{(1)}$ we derived connection between $a^*$, homoatomic interaction energy of Fe and mixing energy to calculate the first nearest neighbour homoatomic interaction energy of Cu, $\epsilon_{\text{CuCu}}^{(1)}$ given as
\begin{equation}
\epsilon_{\text{CuCu}}^{(1)}=\epsilon_{\text{FeFe}}^{(1)} - \frac{2 E_{\text{mixFeCu}}}{2z_1+z_2} a^*
\label{cucuasymmetry}
\end{equation}
In order to calculate interatomic interaction energies experimental thermodynamic data given in Table \ref{podatkiFeCu} were used.
\begin{table}[!ht!bp]
	\caption{Material properties of {Fe} and {Cu} used for calculation of interaction energies and kinetic parameters.}
		\vspace{1mm}
			\begin{center}
				\begin{tabular}{lll}
				\hline
				Cohesive energy of {Fe} & $E_{\text{cohFe}}$ & $-4.28$ eV, \cite{kittel1996} \\
				Mixing energy of {Fe-Cu} & $E_{\text{mixFeCu}}$ & $-$0.515 eV, \cite{salje1833} \\
				Vacancy formation energy of {Fe} & $E_{\text{VforFe}}$ & 1.60 eV, \cite{LBatomic} \\
				Vacancy formation energy of {Cu} & $E_{\text{VforCu}}$ & 1.60 eV, assumption \\
				Vacancy migration energy of {Fe} & $E_{\text{VmigFe}}$ & 0.90 eV, \cite{LBatomic} \\
				Vacancy migration energy of {Cu} in {Fe} & $E_{\text{VmigCu}}$ & 0.70 eV, \cite{LBatomic} \\
				Diffusion constant of {Fe} & D$_{0\_\text{Fe}}$ & 2.01 $10^{-4}$ m$^{2}$s$^{-1}$, \cite{LBdiff} \\
				Attempt frequency of Fe  & $\nu_{\text{Fe}}$ & $8.70\times 10^{12}$ s$^{-1}$, \cite{kittel1996} \\
				Attempt frequency of Cu  & $\nu_{\text{Cu}}$ & $6.67\times 10^{12}$ s$^{-1}$, \cite{kittel1996} \\
				\hline
				\end{tabular}
			\end {center}
		\label{podatkiFeCu}
\end{table}

The ghost interaction energy $\epsilon_{\text{XV}}$ between the solute X and vacancy V is connected to the vacancy formation energy $E_{\text{VforX}}$ and cohesive energy given as	 
\begin{equation}
E_{\text{VforX}} = z_1 \epsilon_{\text{XV}}^{(1)} - E_{\text{cohX}}
\label{vacformation}
\end{equation}
For Fe this can be calculated from experimental thermodynamic data given in Table \ref{podatkiFeCu}. In the case of Cu virtual cohesive energy was used, to calculate ghost interaction energy by reordering equation (\ref{par_cohesive}).

\section{Results and discussion}

To investigate different kinetic pathways leading to same final  microstructure and to elucidate the influence of the asymmetry parameter $a^*$ kMC simulations were performed using a simulation box with 64$^3$ BCC lattices (524288 possible solute positions), with a single vacancy and full periodic boundary conditions in all directions. Since obtained {Cu} precipitates are small, their fully coherency with {Fe} matrix was assumed to justify use of the rigid lattice in simulations. Values of $a^*$ were chosen arbitrary in the range $-10/3$ and $10/3$ and procedure presented in Section II used to determine simulations parameters. As a control run, the kMC simulation with fully symmetric energies ($a^*$=0) was performed. Comparison of interatomic parameters used in our simulations to ones determined by electron calculations like density functional theory (DFT) in same alloy system~\cite{soissonCluster2007} revealed that an asymmetry parameter between 1 and 1.75 should give results comparable to experimental observations.
Simulation box was randomly populated with Cu atoms and quenched from infinite temperature to a temperature $T=873$ K. In order to increase statistics of the cluster properties in the focused regimes of growth and coarsening concentration of Cu was set to 3 at.\% (Cu atoms).
To prevent influence of vacancy trapping in Cu rich areas time was only advanced when vacancy resided in the Fe matrix. The calculated evolution kinetics were described with three different properties, short range order parameter $\alpha(t)$ connected to the pair correlation function, average cluster size defined as second moment of the cluster size distribution $L^{2}(t)$ and by number of atoms binned in clusters larger than pentamers. More detailed description of each property can be found in~\cite{BombacPhD}. Unless stated, all  simulations were run for same time and were stopped after reaching 200k MCS. Asymmetry parameter influences time scales through transition probabilities (cf. equation (\ref{eq:11})). When positive asymmetry was chosen simulations needs much longer runs to reach same times. Due to higher total transition probabilities for each transition, simulation runs for asymmetry parameters $7/3$ and $10/3$ reached small physical times although simulations were let run five and ten times longer and were stopped at 1000k MCS and 2000k MCS, respectively. This happens due to influence of the model asymmetry which changes preferred environment of the vacancy. When $a^*<0$, the vacancy prefers to reside in Fe rich areas and inversely when $a^*>0$ vacancy prefers Cu rich areas which changes kinetic pathways as trapped vacancy in the cluster causes moment of whole cluster.

\begin{figure}[h!t]
\begin{center}
\includegraphics[width=0.90\textwidth]{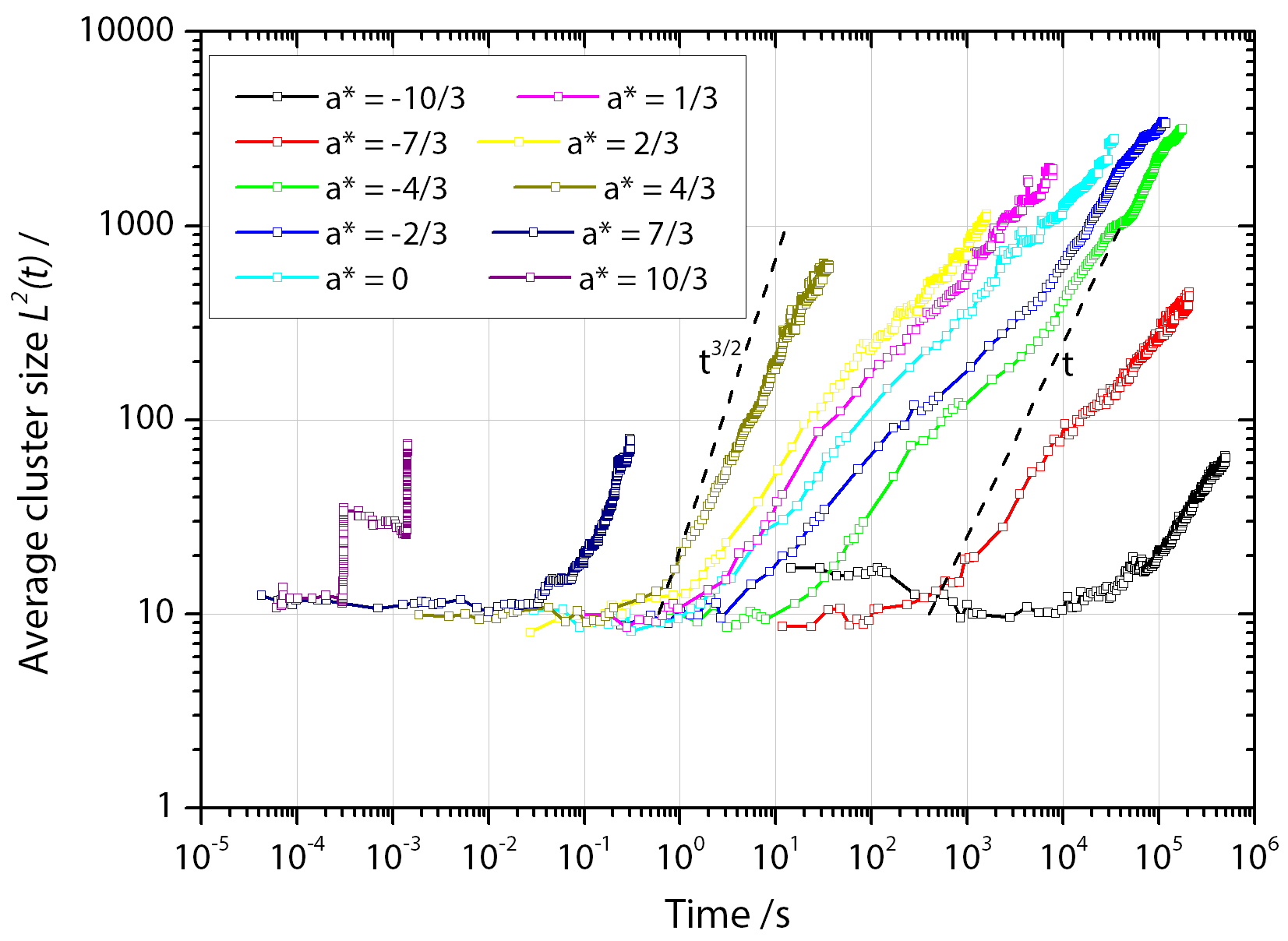}
\caption{Influence of the asymmetry parameter $a^*$ on evolution of the average cluster size.}
\label{L2Asym2}
\end{center}
\end{figure}

Evolution of average cluster size represented as second moment of the cluster size distribution $L^{2}(t)$ is for simulated values of $a^*$ shown in figure \ref{L2Asym2}. In the same figure cluster evolutions corresponding to $t^{3/2}$ and $t$ are depicted with dashed line, representing the classical growth regime and coarsening according to Lifshitz-Slyozov-Wagner (LSW), respectively. Distinctive properties can be observed from the shape of curves. In general, after initial growth stage, cluster growth decreases which is in agreement to classical description. Asymmetry parameter $a^*=1$ behaves similar as description by theoretical growth and coarsening regimes. Both growth and coarsening observed are slower than $t^{3/2}$ and $t$. This is attributed to the change of kinetic pathways and affinity of vacancy and is in agreement to other studies~\cite{FraPRB94, soissonCluster2007}. For $a^*<0$ growth and especially coarsening is diminished since vacancy prefers Fe matrix. In the case of $a^*>0$ vacancy becomes trapped in Cu clusters and changes dynamics of precipitate growth and coarsening. When clusters are small, vacancy trapping does not have a large effect. In the coarsening stage, trapped vacancies can not contribute to the long-range diffusion of Cu in the matrix which is needed for desorption/adsorption process significant for LSW coarsening regime. Coarsening stage for $a^*>0$ happens via different mode through coagulation of clusters as whole. This is confirmed with steps on curves for evolution of mean cluster size (cf. figure \ref{L2Asym2}) where in later stage sudden jump in average size is detected and is attributed to diffusion of whole Cu clusters even large one.

\begin{figure}[hb]
\begin{center}
\includegraphics[width=0.9\textwidth]{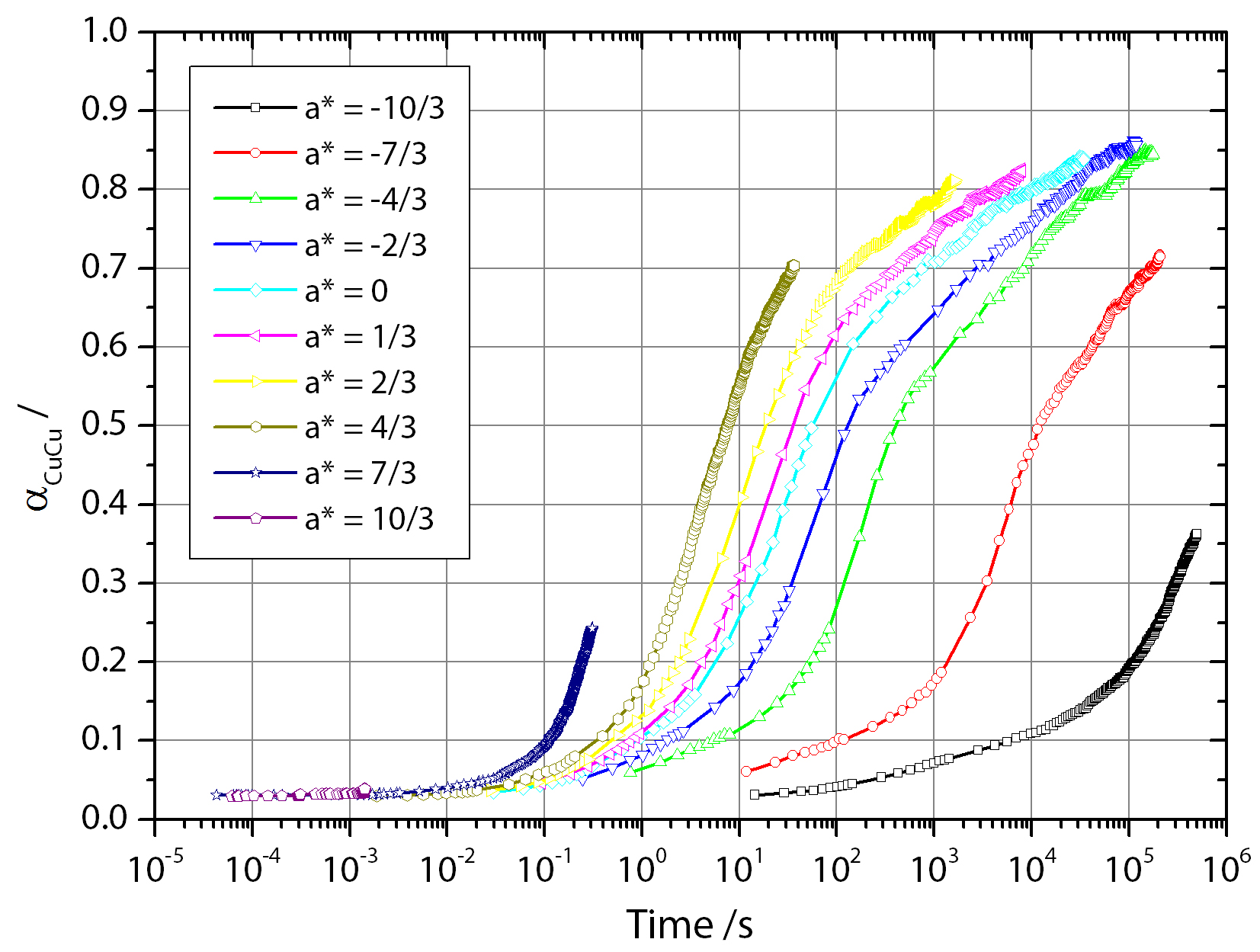}
\caption{Influence of the asymmetry parameter $a^*$ on the evolution of the SRO parameter $\alpha_{\text{CuCu}}^{1}$.}
\label{alfaAsym2}
\end{center}
\end{figure}
Figure \ref{alfaAsym2} depicts the influence of the asymmetry parameter $a^*$ on the evolution of the short range order (SRO) parameter $\alpha_{\text{CuCu}}^{1}$. Parameter $\alpha_{\text{CuCu}}^{1}$ describes the number of Cu atoms paired with another Cu atom. All curves show fast increase of SRO parameter in growth stage and after increase is substantially slowed. Shape of the curve allows identification of different kinetic pathways. For example, when coarsening occurs with LSW mechanism the number of paired Cu atoms will not increase evenly, as smaller clusters need to dissolve so that larger ones can grow causing slowdown in SRO parameter growth. When coarsening occurs via agglomeration of whole clusters, SRO parameter exhibits sudden increase as can be seen in figure \ref{alfaAsym2} for positive $a^*$. In the case of negative $a^*$ SRO parameter enlarges much more evenly, distinctive for desorption/adsorption process. 

\begin{figure}[h!]
\begin{center}
\includegraphics[width=0.9\textwidth]{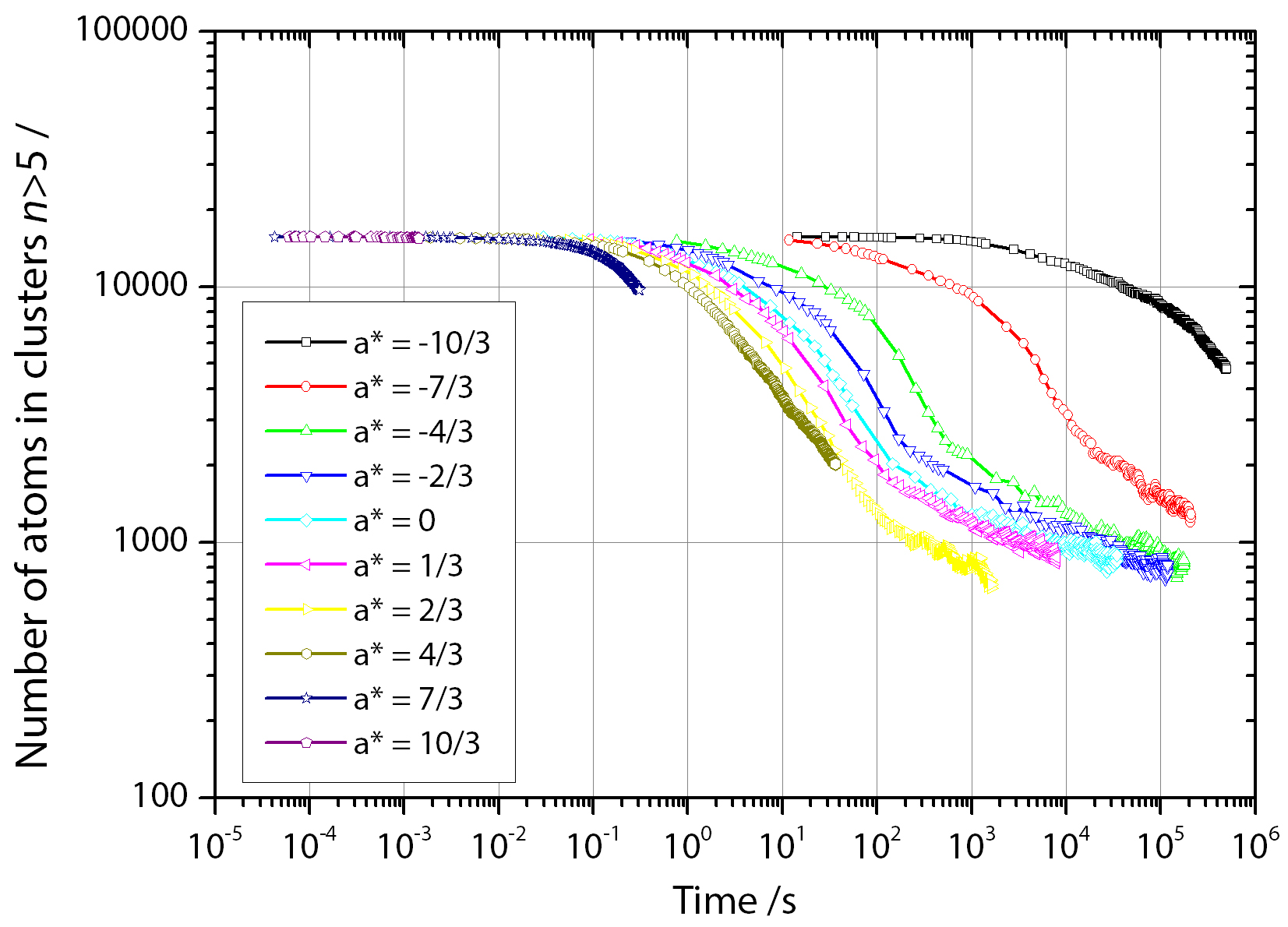}
\caption{Influence the asymmetry parameter $a^*$ on the number of clusters larger than pentamers.}
\label{NtAsym2}
\end{center}
\end{figure}

Different kinetic pathways were also detected by following the number of atoms belonging to clusters. Figure \ref{NtAsym2} shows evolution of atoms in clusters smaller than pentamers. Three distinctive behaviours depending on $a^*$ can be seen. For negative asymmetry parameter number of small clusters is much higher even at longer times. This is attributed to affinity of vacancy which in these case prefers Fe rich matrix and diminishes transport of Cu in the simulation box. Confirmation of this can be seen on the curve for $a^*=-10/3$ in figure \ref{NtAsym2}, where the number of small clusters is very high. Similar to evolution of SRO parameter, sharp border between growth and coarsening stages can be observed. Agglomeration of clusters was observed in later stages  for $a^*>0$. Slope change in the cases where agglomeration is preferred is less steep since clusters are joined together suddenly while in the case of coarsening via LSW regime growth of larger ones happens on expense of smaller. Although smaller clusters are dissolving, they are still present when results are analysed until they become smaller than cut off size set at $n\geq5$.

\begin{figure}
\centering
\subfigure(a){
\includegraphics[width=0.44\textwidth]{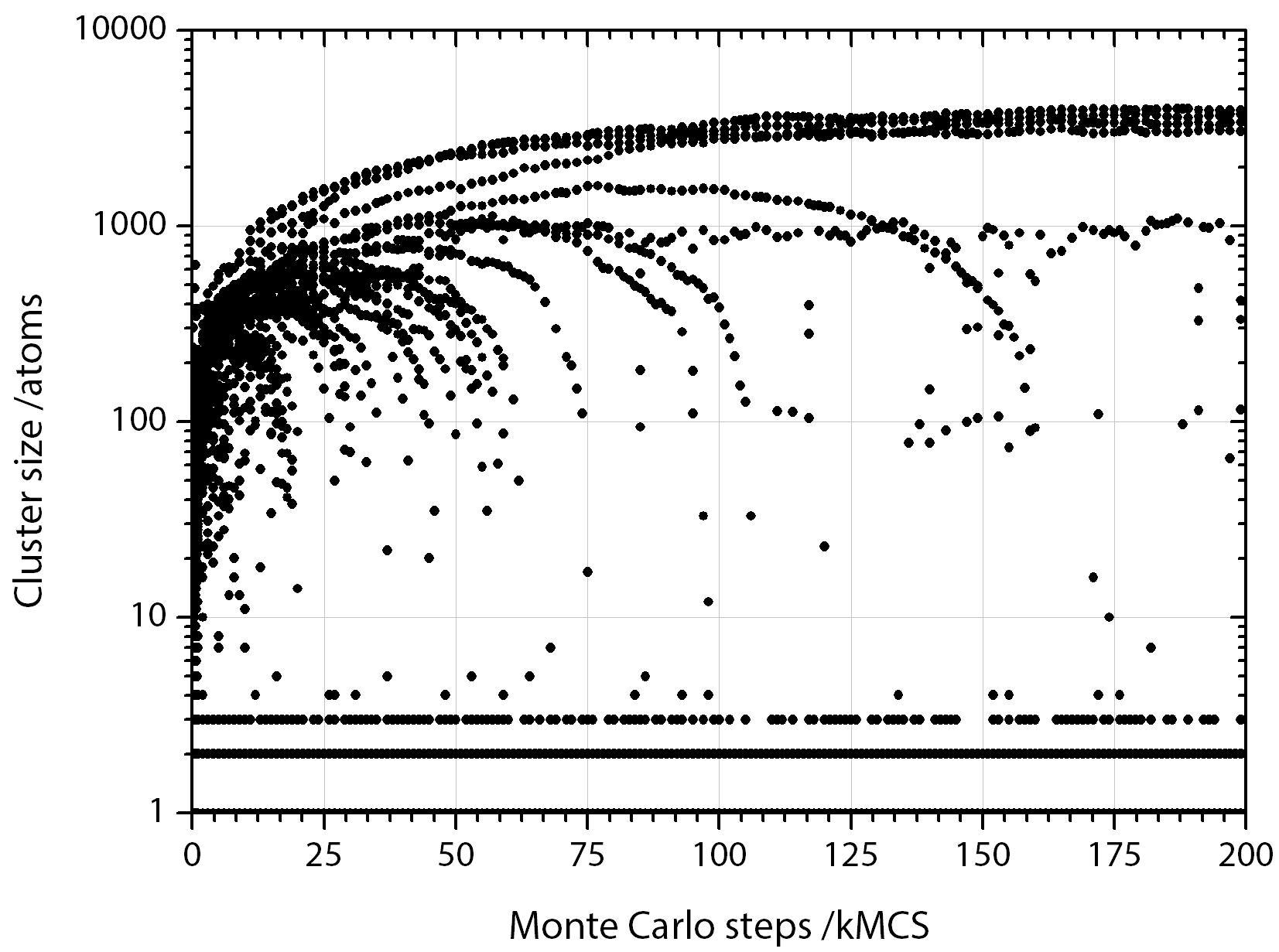}}\
\subfigure(b){
\includegraphics[width=0.44\textwidth]{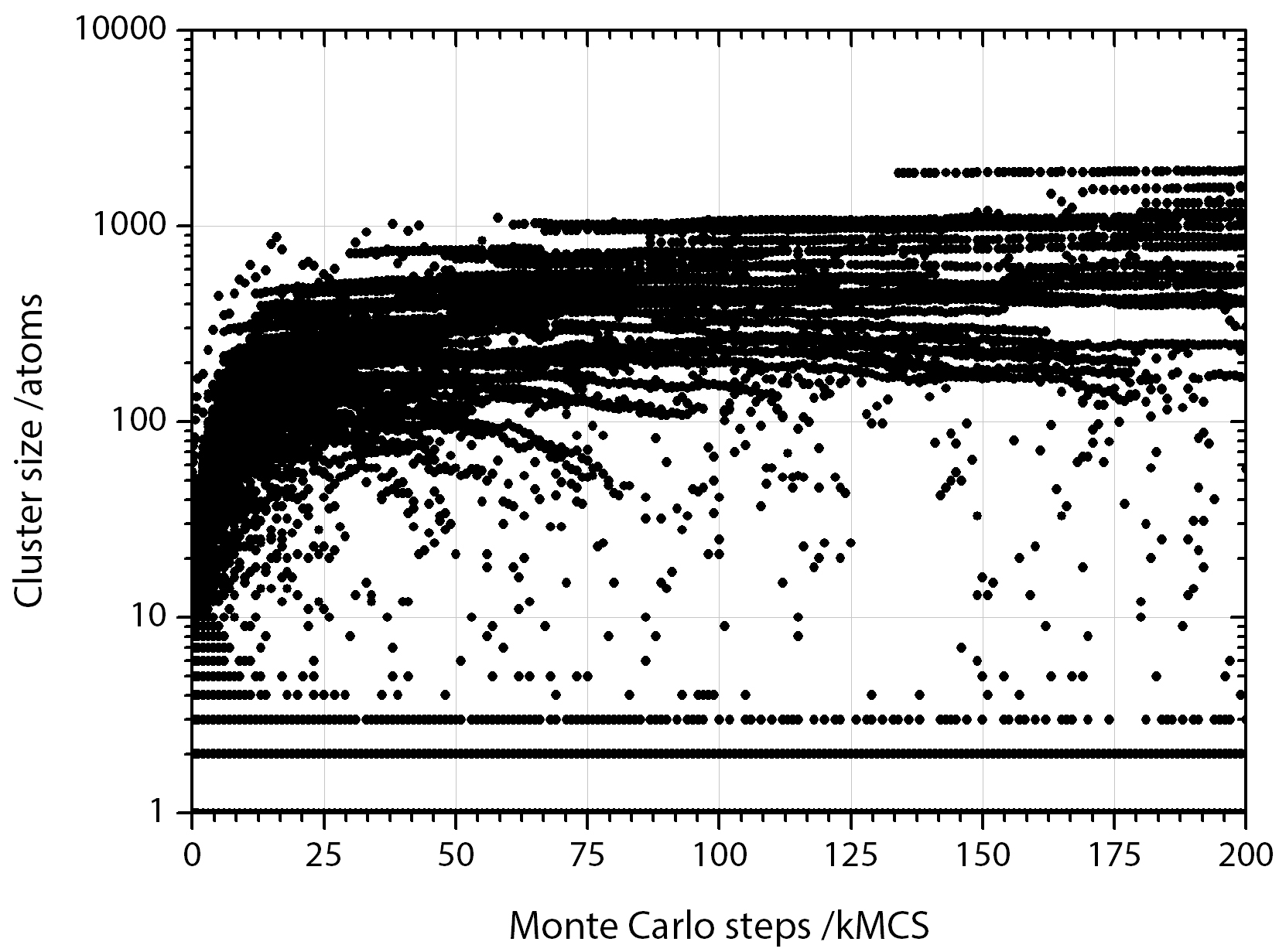}}
\\
\subfigure(c){
\includegraphics[width=0.44\textwidth]{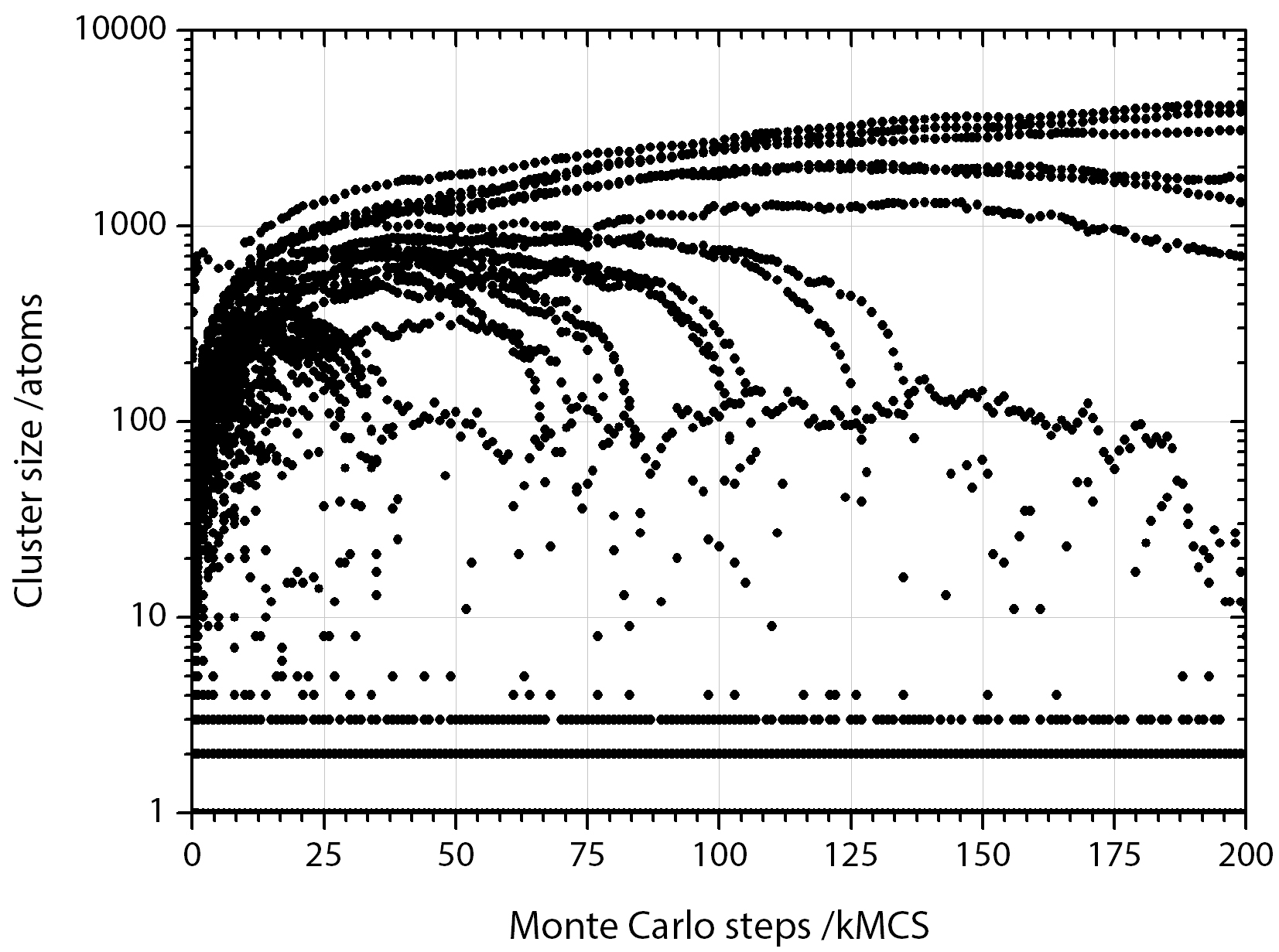}}\
\subfigure(d){
\includegraphics[width=0.44\textwidth]{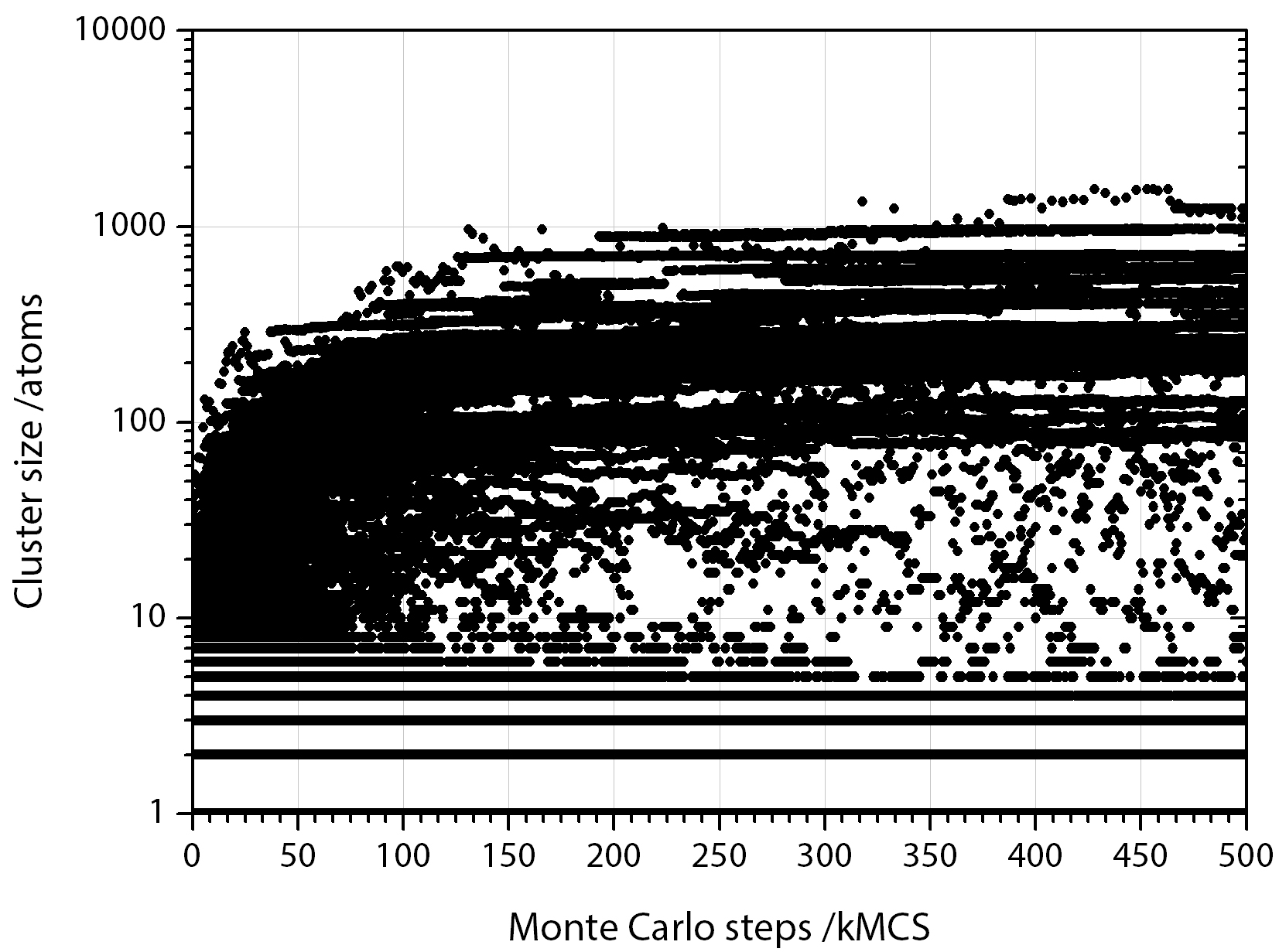}}
\caption{Evolution of clusters displayed in stair-fountain diagram for asymmetry parameters a) $a^*=-2/3$, b) $a^*=2/3$, c) $a^*=-4/3$ and d) $a^*=4/3$.}
\label{fig:fountain}
\end{figure}

In figure \ref{fig:fountain} characteristically different kinetic pathways are shown for asymmetry parameter  $a^*=-2/3$, $a^*=2/3$, $a^*=-4/3$ and $a^*=4/3$ in the form of \emph{stair-fountain} diagrams (cf. reference~\cite{rousselPRB}). In star-fountain diagrams all clusters with Cu atom in adjacent neighbouring position were considered. For negative asymmetry parameter (see figure \ref{fig:fountain}a) evolution of precipitates in later stages resembles a fountain and is happening trough adsorption/desorption as we can follow continues increase and decrease of the number of atoms in large and small clusters, respectively. In the case of positive asymmetry parameter, evolution of clusters depicted in stair-fountain diagram shown in figure \ref{fig:fountain}b exhibits a sudden jump in the number of atoms in clusters, resembling stairs. These jumps are connected with agglomeration of diffusing clusters which collide on their path causing instant increase in the number of atoms in detected clusters. The same also holds for decreased and increased asymmetry parameter depicted in figures \ref{fig:fountain}c and \ref{fig:fountain}d for $a^*=-4/3$ and $a^*=4/3$, respectively.

\begin{figure}[h!]
\begin{center}
\includegraphics[width=0.9\textwidth]{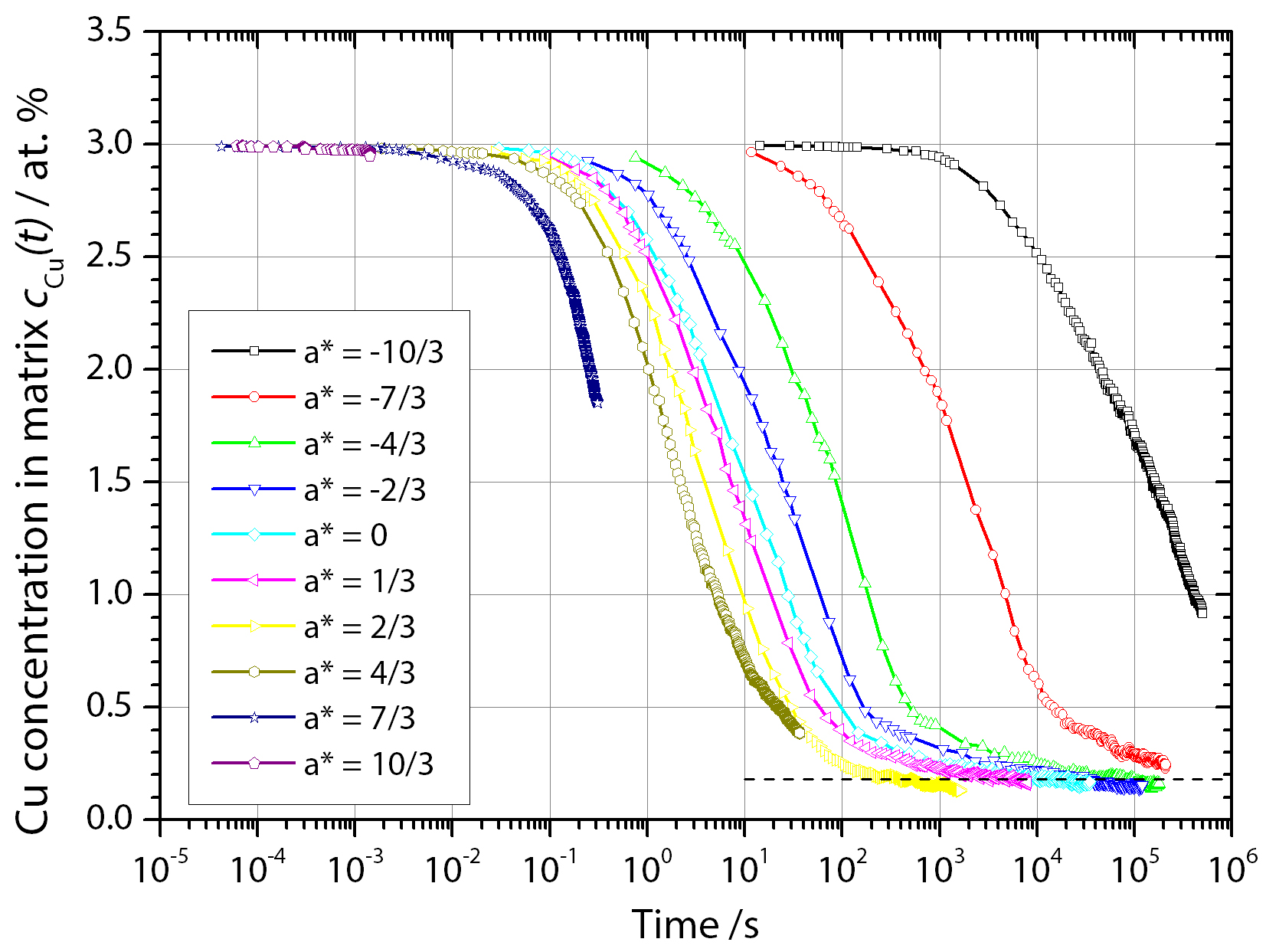}
\caption{Evolution of matrix concentration for various asymmetry parameters $a^*$. Black dashed line represents thermodynamic equilibrium matrix concentration of Cu.}
\label{matrixC2}
\end{center}
\end{figure}

Since $a^*$ has no influence on the resulting phase diagram, the same equilibrium matrix concentration of Cu should be reached in all simulations. In figure \ref{matrixC2} matrix concentration of Cu is shown for all $a^*$used in kMC simulation. In the same figure experimentally determined thermodynamic equilibrium matrix concentration of Cu in Fe at 873 K ($0.18\pm0.03$ at.\%~\cite{Miller1998}) is shown with dashed line.

Obtained results are in agreement with experimental equilibrium matrix concentration for simulations where $a^*>-4/3$. For lower asymmetry parameter (e.g. $a^*=-10/3$) the vacancy prefers Cu rich environment and becomes trapped inside precipitate~\cite{FraPRB94, soissonCluster2007, Warczok201259}. 
Trapped vacancy causes diffusion of cluster as a whole. This moving cluster can on its way meet another cluster, resulting  in agglomeration of both clusters. 
In simulations where $a^*>4/3$, time evolution is very slow and although simulation times were very large equilibrium was not reached. However, comparing evolution curves the convergence towards equilibrium value can be seen. 

\begin{figure}
\centering
\subfigure(a){
\includegraphics[width=0.44\textwidth]{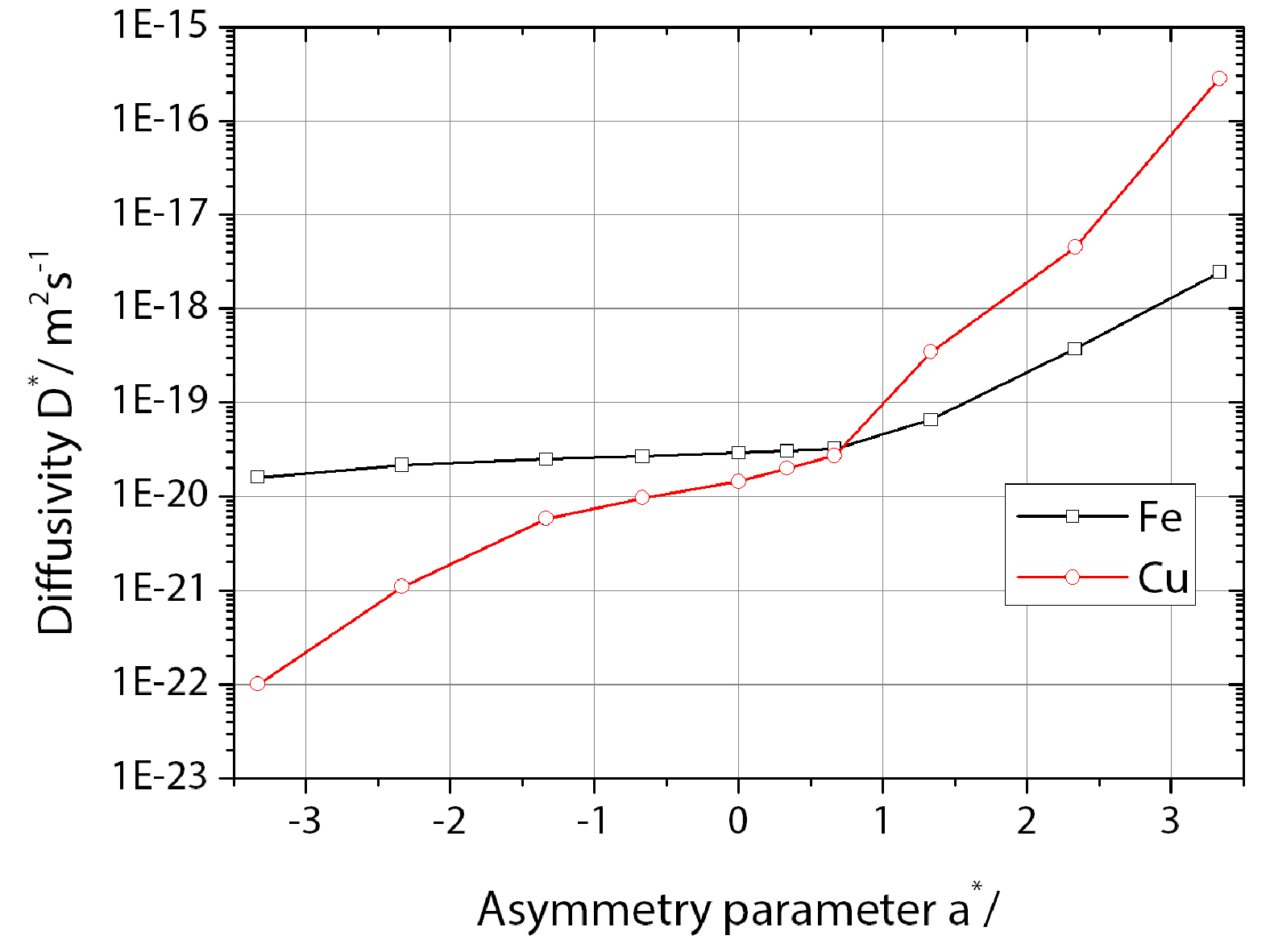}}\
\subfigure(b){
\includegraphics[width=0.44\textwidth]{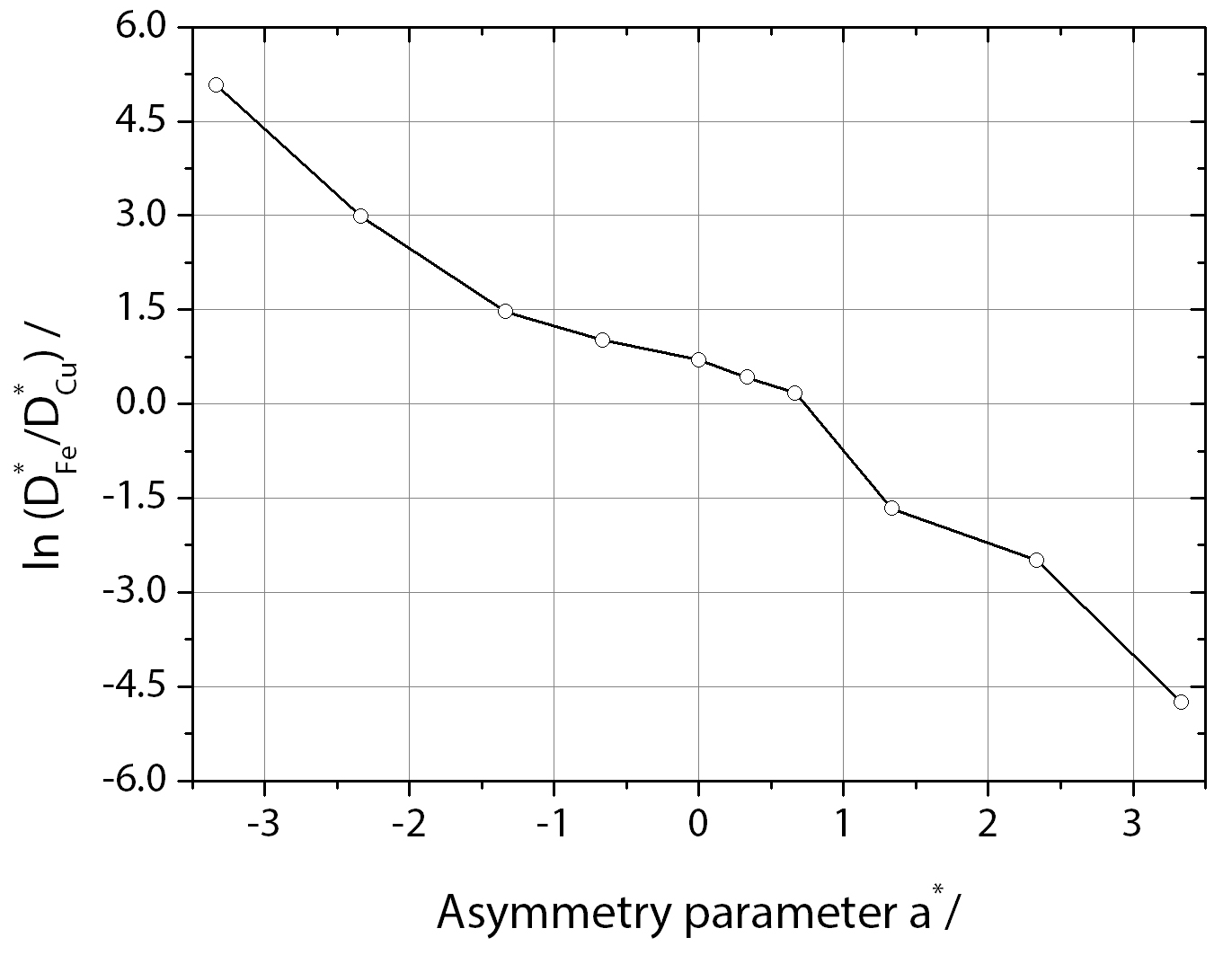}}
\caption{Influence of asymmetry parameter $a^*$ on a) diffusivity of Fe and Cu, and b) logarithm of ratio $D^*_{\rm{Fe}}$/$D^*_{\rm{Cu}}$.}
\label{fig:diff}
\end{figure}

The most convenient result obtained from simulations to compare with experimentally determined are diffusivity of Cu and Fe.
In kMC simulations diffusivity can be calculated from the mean displacement $\left\langle R^2(t)\right\rangle$ of all atoms of each type by
\begin{equation}
D^*=\frac{\left\langle R^2(t)\right\rangle}{2 d t}
\label{tracer}
\end{equation}
where $d$ is dimensionality of the system (3 in our case) and $t$ is time over which mean displacement is followed. In table \ref{Diffusivity} values for Cu and Fe diffusivity at the end of simulations along with ratio between them are presented.
\begin{table}
\caption{Influence of $a^*$ on diffusivity of Cu.}
\vspace{1mm}
	\centering
	\ra{1.2}
		\begin{tabular}{c|ccc}
			\hline
			$a^*$ & $D^*_{\rm{Cu}}$ / m$^2$s$^{-1}$ & $D^*_{\rm{Fe}}$ / m$^2$s$^{-1}$ & $D^*_{\rm{Fe}}$/$D^*_{\rm{Cu}}$ \\
			\hline
			$-10/3$ & $9.99\times10^{-23}$ & $1.592\times10^{-20}$ & 159.4 \\
		  $-7/3$ & $1.09\times10^{-21}$ & $2.16\times10^{-20}$ & 19.7 \\
		  $-4/3$ & $5.81\times10^{-21}$ & $2.51\times10^{-20}$ & 4.32 \\
		  $-2/3$ & $9.64\times10^{-21}$ & $2.66\times10^{-20}$ & 2.76 \\
			0 & $1.44\times10^{-20}$ & $2.91\times10^{-20}$ & 2.02 \\
			1/3 & $1.99\times10^{-20}$ & $3.05\times10^{-20}$ & 1.53 \\
			2/3 & $2.74\times10^{-20}$ & $3.24\times10^{-20}$ & 1.19 \\
			4/3 & $3.47\times10^{-19}$ & $6.56\times10^{-20}$ & 0.19 \\
			7/3 & $4.49\times10^{-18}$ & $3.73\times10^{-19}$ & 0.08 \\
			10/3 & $2.82\times10^{-16}$ & $2.44\times10^{-18}$ & 0.01 \\
				\hline
		\end{tabular}
		\label{Diffusivity}
\end{table}
Graphical representation of data in table \ref{Diffusivity} is given in figure \ref{fig:diff}. Figure \ref{fig:diff}a depicts influence of asymmetry parameter $a^*$ on diffusivity of Fe and Cu where it is shown that diffusivity of Fe is not influenced by decreased in asymmetry parameter. When asymmetry parameter is increased diffusivity of Fe starts to rise, however rise is much smaller in comparison to Cu. For highly negative $a^*$ explained with preferred environment for vacancy in Fe and for highly positive $a^*$ by trapping of vacancy in Cu cluster and diminished mean distance of Cu atoms. From this we can conclude, that significant changes in diffusivity of Cu are due to preferred area of vacancy. Ratio $D^*_{\rm{Fe}}$/$D^*_{\rm{Cu}}$ is for values of $a^*$ simulated, shown in figure \ref{fig:diff}b where can be seen that ratio exhibits decrease if asymmetry parameter is increased.

Salje and Feller-Kniepmeier~\cite{salje1833} have experimentally determined equation for diffusivity of Cu in $\alpha$-Fe 
\begin{equation}
D^{\alpha-{\rm{Fe}}}_{\rm{Cu}}(T)  \ /{\rm{cm}}^2{\rm{s}}^{-1} = 0.63 \exp\left(-\frac{2.29}{k_B T} \right)
\label{saljeEQ}
\end{equation}
At 873 K equation (\ref{saljeEQ}) gives diffusivity of Cu in $\alpha$-Fe is $D^{\alpha-{\rm{Fe}}}_{\rm{Cu}}=3.8\times10^{-18}$ m$^2$s$^{-1}$ and is similar to results obtained by other authors where $D^{\alpha-{\rm{Fe}}}_{\rm{Cu}}=1.76\times10^{-18}$ m$^2$s$^{-1}$~\cite{Golubov2000}. 

Furthermore, experiments have confirmed that diffusion of Cu in $\alpha$-Fe is faster than self-diffusion of Fe~\cite{LBdiff}. Results obtained from kMC simulations presented in table \ref{Diffusivity} that correspond to this fact are the ones with asymmetry parameter $a^* \approx 4/3$ and more. Furthermore, kMC results obtained with $a^* = 4/3$ are also in agreement with experimentally determined diffusivity~\cite{salje1833}. Using asymmetry $a^* = 4/3$, value for corresponding virtual cohesive energy of Cu is $E^{\rm{virtual}}_{\text{cohCu}}=-3.94$ eV and is in agreement with previous kMC results~\cite{LeBou, soissonCluster2007, Vincent2008}. 

The presented model can be used to quickly obtain vacancy diffusion parameters dependent on asymmetry and perform kMC simulations of realistic behaviour without the need for expensive electronic calculations of interaction parameters. Although the parametrization model was developed and presented on binary alloy it can be expanded to arbitrary number of interaction solutes. In the case of more than two interacting elements, the number of asymmetry parameters increases, for example in the case of ternary ABC system number of possible asymmetry parameters increases to three with $a^*_{\rm{AB}}$, $a^*_{\rm{AC}}$ and $a^*_{\rm{BC}}$.

\section{Conclusions}

In this paper a simple model based on thermodynamics with an asymmetry of interatomic potentials is presented. This allows quick determination of simulation parameters without the need for expensive electron calculations. Its usage is demonstrated by kinetic Monte Carlo simulations, where various kinetic pathways leading to the same final microstructure were explored on model Fe-Cu alloy. Kinetic pathways are controlled through interatomic interaction energies defined with help of asymmetry parameter. Simulations revealed  that for positive asymmetry parameter coarsening through agglomeration of clusters dominated over Ostwald ripening mechanism. For symmetrical model and negative asymmetry parameter growth and coarsening happened through classical description. Kinetic pathways for model with large negative asymmetry parameter are substantially changed by affinity of vacancy to reside in Fe matrix. Comparison to available experimental diffusion data reveals that behaviour of real Fe-Cu alloy is in agreement with results obtained with asymmetry parameter $a^*=4/3$.

\section*{Acknowledgements}

This work was supported by the Slovenian Research Agency (ARRS) under grant 1000-06-310075.



\appendix
\section{Parameters used for simulations}\label{param}
Interatomic interaction energies used for kMC simulations were calculated using procedure presented in Section II. In table \ref{asymmetrypars} all used parameters for particular arbitrary chosen $a^*$ are given. Interaction energies for Fe-Fe and Fe-V were same in all simulations and values used are $\epsilon_{\text{FeFe}}^{(1)} = -0.77818$ eV, $\epsilon_{\text{FeFe}}^{(2)} = -0.38909$ eV and $\epsilon_{\text{FeV}}^{(1)} = -0.335$ eV.
\begin{sidewaystable}[!htbp]
 \ra{1.2}
	\caption{Interaction parameters used for kMC simulations.}
		\vspace{1mm}
			\begin{center}
				\begin{tabular}{c|cccccccccc}
				\hline 
				Binding & \multicolumn{9}{|c}{Asymmetry parameter $a^*$}  \\
				energy /eV  & -10/3 & -7/3 & -4/3 & -2/3 & 0 & 1/3 & 2/3 & 4/3 & 7/3 & 10/3 \\
				\hline
				$\epsilon_{\text{CuCu}}^{(1)}$ & -0.93424 & -0.88742 & -0.84061 & -0.80939 & -0.77818 & -0.76258 & -0.74697 & -0.71576 & -0.66894 & -0.62212 \\
				$\epsilon_{\text{CuCu}}^{(2)}$ & -0.46712 & -0.44371 & -0.4203 & -0.4047 & -0.38909 & -0.38129 & -0.37348 & -0.35788 & -0.33447 & -0.31106 \\
				$\epsilon_{\text{FeCu}}^{(1)}$ & -0.80939 & -0.78598 & -0.76258 & -0.74697 & .0,73136 & -0.72356 & -0.71576 & -0.70015 & -0.67674 & -0.65333 \\
				$\epsilon_{\text{FeCu}}^{(2)}$ & -0.4047 & -0.39299 & -0.38129 & -0.37348 & -0.36568 & -0.36178 & -0.35788 & -0.35008 & -0.33837 & -0.32667 \\
				$\epsilon_{\text{CuV}}^{(1)}$  & -0.44229 & -0.4101 & -0.37792 & -0.35646 & -0.335 & -0.32427 & -0.31354 & -0.29208 & -0.2599 & -0.22771 \\
				\hline
				\end{tabular}
			\end {center}
		\label{asymmetrypars}
\end{sidewaystable}

\newpage

\section*{References}



\bibliographystyle{unsrt}
\bibliography{references}

\begin{thebibliography}{10}

\bibitem{christian}
J.~W. Christian.
\newblock {\em The Theory of Transformations in Metals and Alloys}.
\newblock Pergamon, 3$^{\text{rd}}$ edition, 2002.

\bibitem{young}
W.~M. Young and E.~W. Elcock.
\newblock Monte {C}arlo studies of vacancy migration in binary ordered alloys:
  {I}.
\newblock {\em Proceedings of the Physical Society}, 89:735, 1966.

\bibitem{SoisAM96}
F.~Soisson, A.~Barbu, and G.~Martin.
\newblock Monte {C}arlo simulations of copper precipitation in dilute
  iron-copper alloys during thermal ageing and under electron irradiation.
\newblock {\em Acta Materialia}, 44(9):3789--3800, 1996.

\bibitem{AthAM96}
M.~Ath{\`e}nes, P.~Bellon, G.~Martin, and F.~Haider.
\newblock A {M}onte-{C}arlo study of {B}2 ordering and precipitation via
  vacancy mechanism in b.c.c. lattices.
\newblock {\em Acta Materialia}, 44(12):4739--4748, 1996.

\bibitem{schmauder}
S.~Schmauder and P.~Binkele.
\newblock Atomistic computer simulation of the formation of {C}u-precipitates
  in steels.
\newblock {\em Computational Materials Science}, 24:42--53, 2002.

\bibitem{cerezo}
G.~Sha and A.~Cerezo.
\newblock Kinetic {M}onte {C}arlo simulation of clustering in an {Al–Zn–Mg–Cu}
  alloy (7050).
\newblock {\em Acta Materialia}, 53(4):907--917, February 2005.

\bibitem{Mao2011}
Z.~Mao, C.~Booth-Morrison, C.~K. Sudbrack, G.~Martin, and D.~N. Seidman.
\newblock Kinetic pathways for phase separation: {A}n atomic-scale study in
  {N}i-{A}l-{C}r alloys.
\newblock {\em Acta Materialia}, 60(4):1871--1888, February 2012.

\bibitem{BombacPhD}
D.~Bombac.
\newblock {\em Atomistic Simulations of Precipitation Kinetics in
  Multicomponent Interstitial/Substitutional Alloys}.
\newblock PhD thesis, University of Ljubljana, 2012.

\bibitem{Warczok201259}
Piotr Warczok, Jaroslav \v{Z}en{\'i}\v{s}ek, and Ernst Kozeschnik.
\newblock Atomistic and continuums modeling of cluster migration and
  coagulation in precipitation reactions.
\newblock {\em Computational Materials Science}, 60(0):59--65, 2012.

\bibitem{Osetsky1994236}
Yu.N. Osetsky, A.G. Mikhin, and A.~Serra.
\newblock Computer simulation study of copper precipitates in $\alpha$-iron.
\newblock {\em Journal of Nuclear Materials}, 212-215, Part 1(0):236--240,
  1994.

\bibitem{Nagano2006223}
Takatoshi Nagano and Masato Enomoto.
\newblock Simulation of the growth of copper critical nucleus in dilute bcc
  {Fe-Cu} alloys.
\newblock {\em Scripta Materialia}, 55(3):223--226, 2006.

\bibitem{Deschamps2010236}
Alexis Deschamps and Michel Perez.
\newblock Mesoscopic modelling of precipitation: {A} tool for extracting
  physical parameters of phase transformations in metallic alloys.
\newblock {\em Comptes Rendus Physique}, 11(3-4):236--244, 2010.
\newblock Computational metallurgy and scale transitions M{\~A}©tallurgie
  num{\~A}©rique et changements d'{\~A}©chelle.

\bibitem{othenPM}
P.~J. Othen, M.~L. Jenkins, and G.~D.~W. Smith.
\newblock High-resolution electron microscopy studies of the structure of {C}u
  precipitates in $\alpha$-{F}e.
\newblock {\em Philosophical Magazine A}, 70(1):1--24, 1994.

\bibitem{pizzini}
S.~Pizzini, K.~J. Roberts, W.~J. Phythian, C.~A. English, and G.~N. Greaves.
\newblock A fluorescence {EXAFS} study of the structure of copper-rich
  precipitates in {F}e-{C}u and {F}e-{C}u-{N}i alloys.
\newblock {\em Philosophical Magazine Letters}, 61(4):223--229, 1990.

\bibitem{othenPML}
P.~J. Othen, M.~L. Jenkins, G.~D.~W. Smith, and W.~J. Phythian.
\newblock Transmission electron microscope investigations of the structure of
  copper precipitates in thermally-aged {F}e-{C}u and {F}e-{C}u-{N}i.
\newblock {\em Philosophical Magazine Letters}, 64(6):383--391, 1991.

\bibitem{hrtem95}
H.~A.~Hardouin Duparc, R.~C. Doole, M.~L. Jenkins, and A.~Barbu.
\newblock A high-resolution electron microscopy study of copper precipitation
  in {F}e-1.5 wt\% {C}u under electron irradiation.
\newblock {\em Philosophical Magazine Letters}, 71(6):325--333, 1995.

\bibitem{salje1833}
G.~Salje and M.~Feller-Kniepmeier.
\newblock The diffusion and solubility of copper in iron.
\newblock {\em Journal of Applied Physics}, 48(5):1833--1839, 1977.

\bibitem{LBdiff}
H.~Mehrer, editor.
\newblock {\em Diffusion in {S}olid {M}etals and {A}lloys}, volume~26 of {\em
  Landolt-B{\"o}rnstein}.
\newblock Springer-Verlag, 1991.

\bibitem{monzenPM}
R.~Monzen, M.~L. Jenkins, and A.~P. Sutton.
\newblock The bcc-to-9{R} martensitic transformation of {C}u precipitates and
  the relaxation process of elastic strains in an {F}e-{C}u alloy.
\newblock {\em Philosophical Magazine A}, 80(3):711--723, 2000.

\bibitem{apfecu}
M.~Schober, E.~Eidenberger, P.~Staron, and H.~Leitner.
\newblock Critical {C}onsideration of {P}recipitate {A}nalysis of {F}e -1 at.\%
  {C}u {U}sing {A}tom {P}robe and {S}mall-{A}ngle {N}eutron {S}cattering.
\newblock {\em Microscopy and Microanalysis}, 17:26--33, 2011.

\bibitem{Golubov2000}
S.I. Golubov, A.~Serra, Yu.N. Osetsky, and A.V. Barashev.
\newblock On the validity of the cluster model to describe the evolution of cu
  precipitates in fe{\^a}€“cu alloys.
\newblock {\em Journal of Nuclear Materials}, 277(1):113--115, 2000.

\bibitem{Takahashi2011}
J.~Takahashi, K.~Kawakami, and Y.~Kobayashi.
\newblock Consideration of particle-strengthening mechanism of
  copper-precipitation-strengthened steels by atom probe tomography analysis.
\newblock {\em Materials Science and Engineering: A}, 535:144--152, February
  2012.

\bibitem{FraPRB94}
P.~Fratzl and O.~Penrose.
\newblock Kinetics of spinodal decomposition in the {I}sing model with vacancy
  diffusion.
\newblock {\em Physical Review B}, 50:3477--3480, Aug 1994.

\bibitem{athenesPM97}
M.~Ath{\`e}nes, P.~Bellon, and G.~Martin.
\newblock Identification of novel diffusion cycles in {B}2 ordered phases by
  {M}onte {C}arlo simulation.
\newblock {\em Philosophical Magazine A}, 76(3):565--585, 1997.

\bibitem{athenesPM99}
M.~Ath{\`e}nes and P.~Bellon.
\newblock Antisite-assisted diffusion in the {L}12 ordered structure studied by
  {M}onte {C}arlo simulations.
\newblock {\em Philosophical Magazine A}, 79(9):2243--2257, 1999.

\bibitem{AthAM00}
M.~Ath{\`e}nes, P.~Bellon, and G.~Martin.
\newblock Effects of atomic mobilities on phase separation kinetics: a
  {M}onte-{C}arlo study.
\newblock {\em Acta Materialia}, 48(10):2675--2688, 2000.

\bibitem{rousselPRB}
J.-M. Roussel and P.~Bellon.
\newblock Vacancy-assisted phase separation with asymmetric atomic mobility:
  {C}oarsening rates, precipitate composition, and morphology.
\newblock {\em Physical Review B}, 63:184114, Apr 2001.

\bibitem{soissonCluster2007}
F.~Soisson and C.~C. Fu.
\newblock Cu-precipitation kinetics in $\alpha{}$-{F}e from atomistic
  simulations: {V}acancy-trapping effects and {C}u-cluster mobility.
\newblock {\em Physical Review B}, 76(21):214102, Dec 2007.

\bibitem{Erdelyi20105639}
Z.~Erd{\'e}lyi, Z.~Balogh, and D.L. Beke.
\newblock Kinetic critical radius in nucleation and growth processes --
  {T}rapping effect.
\newblock {\em Acta Materialia}, 58(17):5639--5645, 2010.

\bibitem{Bortz75}
A.~B. Bortz, M.~H. Kalos, and J.~L. Lebowitz.
\newblock A new algorithm for {M}onte {C}arlo simulation of {I}sing spin
  systems.
\newblock {\em Journal of Computational Physics}, 17(1):10--18, 1975.

\bibitem{gillespie}
D.~T. Gillespie.
\newblock A general method for numerically simulating the stochastic time
  evolution of coupled chemical reactions.
\newblock {\em Journal of Computational Physics}, 22:403--434, 1976.

\bibitem{LeBou}
Y.~Le{\phantom{--}}Bouar and F.~Soisson.
\newblock Kinetic pathways from embedded-atom-method potentials: {I}nfluence of
  the activation barriers.
\newblock {\em Physical Review B}, 65(9):94--103, 2002.

\bibitem{ducastelle1991order}
F.~Ducastelle.
\newblock {\em Order and phase stability in alloys}.
\newblock Cohesion and structure. North-Holland, Amsterdam, 1991.

\bibitem{kittel1996}
C.~Kittel.
\newblock {\em Introduction to solid state physics}.
\newblock Wiley, 1996.

\bibitem{LBatomic}
H.~Ullmaier, editor.
\newblock {\em Atomic {D}efects in {M}etels}, volume~25 of {\em
  Landolt-B{\"o}rnstein}.
\newblock Springer-Verlag, 1991.

\bibitem{Miller1998}
M.~K. Miller, K.~F. Russell, P.~Pareige, M.~J. Starink, and R.~C. Thomson.
\newblock Low temperature copper solubilities in {Fe{\^a}€“Cu{\^a}€“Ni}.
\newblock {\em Materials Science and Engineering: A}, 250(1):49--54, 1998.

\bibitem{Vincent2008}
E.~Vincent, C.S. Becquart, C.~Pareige, P.~Pareige, and C.~Domain.
\newblock Precipitation of the fecu system: A critical review of atomic kinetic
  monte carlo simulations.
\newblock {\em Journal of Nuclear Materials}, 373(1-3):387--401, 2008.

\end{thebibliography}







\end{document}